\documentclass[prb,twocolumn,showpacs,floatfix,aps,superscriptaddress]{revtex4}
\usepackage{amssymb}
\usepackage{amsmath}
\usepackage[dvips]{graphicx}
\usepackage{mathrsfs}
\usepackage{ulem}
\usepackage[mathscr]{euscript}
\usepackage{gensymb}
\usepackage{color}
\begin{document}

\title{Gate voltage controlled electronic transport through a
ferromagnet/normal/ferromagnet junction on the surface of a topological
insulator}
\author{Kun-Hua Zhang}
\author{Zheng-Chuan Wang}
\author{Qing-Rong Zheng}
\email{qrzheng@ucas.ac.cn}
\author{Gang Su}
\email{gsu@ucas.ac.cn}
\affiliation{Theoretical Condensed Matter
Physics and Computational Materials Physics Laboratory, School of
Physics, University of Chinese Academy of
Sciences, Beijing 100049, China}

\begin{abstract}
We investigate the electronic transport properties of a
ferromagnet/normal/ferromagnet junction on the surface of a
topological insulator with a gate voltage exerted on the normal
segment. It is found that the conductance oscillates with the width
of normal segment and gate voltage, and the maximum of conductance
gradually decreases while the minimum of conductance approaches zero
as the width increases. The conductance can be controlled by tuning
the gate voltage like a spin field-effect transistor. It is
found that the
magnetoresistance ratio can be very large, and
can also be negative owing to the anomalous transport. In addition,
when there exists a magnetization component in the surface plane, it
is shown that only the component parallel to the junction interface
has an influence on the conductance.
\end{abstract}

\maketitle

\section{Introduction}

Topological insulators are new quantum states discovered recently, which
have a bulk band gap and gapless edge states or metallic surface states due to the
time-reversal-symmetry and spin-orbit coupling interaction~\cite{M.Z.Hasan2010_X.L.Qi2011}.
The two-dimensional (2D)
topological insulator has first been predicted theoretically as a quantum
spin Hall state\cite{C.L.Kane2005,B.A.Bernevig2006}
and then observed experimentally\cite{M.Konig2007}.
The topological characterization of quantum spin Hall insulators can be
generalized from 2D to three-dimensional (3D) case, and leads to the discovery of 3D topological
insulator (TI)\cite{L.Fu2007a,R.Roy2009_J.E.Moore2007,H.Zhang2009,L.Fu2007b}.
The TIs in 3D are usually classified according to the
number of Dirac cones on their surfaces. Those strong topological insulators with odd number of Dirac
cones on their surfaces are robust against the time-reversal invariant disorder,
while the weak topological insulator is referred to those
with even number Dirac cones on their surfaces which depends on the
surface direction and might be broken even without breaking the time
reversal symmetry\cite{L.Fu2007a,L.Fu2007b}. When the TIs are coated with magnetic or
superconducting layers, the surface states could be gapped and many
interesting properties emerge, such as half-integer quantum Hall effect\cite{Y.Zheng2002}
, Majorana fermion\cite{L.Fu2008}, etc.

The topological surface states had been observed by several experimental groups
by means of angle-resolved photoemission spectroscopy (ARPES)\cite{D.Hsieh2008,Y.Xia2009,Y.L.Chen2010}
and scanning tunneling microscopy (STM)\cite{T.Zhang2009,J.Seo2010}. Although the residual bulk
carrier density brings much difficulty to the surface states transport
experiments\cite{J.G.Analytis2010,K.Eto2010}, the signatures of negligible bulk carriers
contributing to the transport\cite{C.Brune2011} and near $100\%$ surface transport in
topological insulator\cite{B.Xia2012} have been found recently in experiments.

The low energy physics of the surface states of strong topological
insulators can be described by the 2D massless Dirac theory\cite{H.Zhang2009}, which is
different from that in graphene where the spinors are composed of different
sublattices\cite{C.W.J.Beenakker2008_A.H.Castro Neto2009}. The topological surface states show strong spin-orbit
coupling, which may be applied to the spin field-effect
transistors in spintronics\cite{S.Datta1990,I.Zutic2004,S.A.Wolf2001,A.Fert2008,Z.G.Zhu2003_B.Jin2003,
H.F.Mu2005_H.F.Mu2006_X.Chen2008}. The electronic transport properties on
topological insulator surface with magnetization has attracted a lot of
attention\cite{T.Yokoyama2010,B.Soodchomshom2010,M.Salehi2011,S.Mondal2010,J.P.Zhang2012,JinhuaGao2009,
Jian-HuiYuan2012,Z.Wu2010_Y.Zhang2010}.
In Refs.~\onlinecite{T.Yokoyama2010} and \onlinecite{B.Soodchomshom2010} the results are given in the
limit of thin barrier (i.e., the width of barrier $L$$\rightarrow $0 and
barrier potential $V_{0}$$\rightarrow \infty $ while $V_{0}L$ is constant),
and the physical origin of this thin barrier is the mismatch effect and built-in
electric field of junction interface. Refs.~\onlinecite{M.Salehi2011}
and \onlinecite{Jian-HuiYuan2012}
studied the spin valve on the surface of topological insulator, in which the exchange fields in the two ferromagnetic leads are assumed to align along the y axis direction. Refs.~\onlinecite{S.Mondal2010},
\onlinecite{JinhuaGao2009}, \onlinecite{J.P.Zhang2012} and \onlinecite{Z.Wu2010_Y.Zhang2010}
investigated the electron transport through ferromagnetic barrier on the surface of a topological insulator. It is noted that both the electric potential barrier and the ferromagnetic barrier are the transport channels in these models.
The bulk band gap of topological
insulator is usually about $20$-$300$ meV\cite{H.Zhang2009,D.Hsieh2008,Y.Xia2009,Y.L.Chen2010,C.Brune2011},
in order to keep the
transport at the Fermi energy inside the bulk gap, and the gate voltage on
topological insulator should be finite.

In this paper, we study the electronic transport through a 2D
ferromagnet/normal/ferromagnet junction on the surface of a strong
topological insulator where a gate voltage is exerted on the normal
segment with a finite width, and the exchange
fields in the two ferromagnetic leads point mainly to the z axis direction.
So far such a system has not been well
studied. We find that the conductance oscillates with the width of
normal segment and gate voltage, and the maximum of conductance
gradually decreases while the minimum of conductance can approach
zero as the width increases. These behaviors are more obvious when
the gate voltage is smaller than the Fermi energy. This
gate-controlled 2D topological ferromagnet/normal/ferromagnet
junction shows the property of a spin field-effect transistor. The
magnetoresistance (MR) can be very large and could also be negative
owing to the anomalous transport. In addition, when there exists a
magnetization component in the 2D plane, it is shown that only the
magnetization component which is parallel to the junction interface
has an influence on the conductance.

This paper is organized as follows. First, we will describe the theoretical model
for the electronic transport through the topological spin-valve junction.
Second, we will present our numerical results and discussions. Finally, a brief summary will be given.

\section{Theoretical Formalism}

\begin{figure}[tbp]
\includegraphics[width=8.5cm]{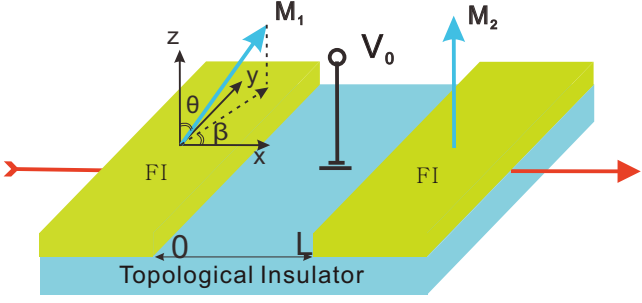}
\caption{(Color online) Schematic layout of a 2D ferromagnet/normal/ferromagnet junction
on the surface of a topological insulator. An exchange split on the surface
underneath the ferromagnetic insulator (FI) is induced by the proximity effect,
and the central normal segment is tuned by a gate voltage $V_{0}$. The current
flows along the x-axis on the surface.}
\label{schematics of FM/Normal/FM junction}
\end{figure}

We consider a 2D ferromagnet/normal/ferromagnet junction on a strong topological
insulator surface as shown in Fig.~\ref{schematics of FM/Normal/FM junction}.
The bulk ferromagnetic insulator (FI) interacts with
the surface electrons in TI by the proximity effect, and the ferromagnetism is induced in the
topological surface states\cite{T.Yokoyama2010,B.Soodchomshom2010,M.Salehi2011,S.Mondal2010,J.P.Zhang2012,
Z.Wu2010_Y.Zhang2010,H.Haugen2008,I.Vobornik2011,Weidong Luo2012}. The interfaces between ferromagnet (FM) and normal
segment are parallel to $y$ direction, and the normal segment is located between $x=0$ and $x=L$ with
gate voltage $V_{0}$ exerted on it\cite{H.Steinberg2011,Y.Wang2012,J.R.Williams2007}. Here we presume, for the simplicity, the distance $L$
between two interfaces is shorter than the mean free path as well as the spin coherence length.
With this setup, the Hamiltonian for this system reads\cite{T.Yokoyama2010,B.Soodchomshom2010,M.Salehi2011,S.Mondal2010,J.P.Zhang2012,
Z.Wu2010_Y.Zhang2010}
\begin{equation}
\widehat{H}= \upsilon _{F}\widehat{\sigma}\cdot \widehat{p}+ \widehat{\sigma}%
\cdot \overset{\rightharpoonup}{m}(r) +V(r)
\label{Hamiltonian}
\end{equation}
with Pauli matrices $\widehat{\sigma}=(\widehat{\sigma}_{x},\widehat{\sigma}%
_{y},\widehat{\sigma}_{z})$, the in-plane electron momentum $\widehat{p}=(%
\widehat{p}_{x},\widehat{p}_{y},0)$, and Fermi velocity $\upsilon _{F}$. The
piecewise magnetization $\overset{\rightharpoonup}{m}(r)$ is chosen to be a 3D
vector pointing along an arbitrary direction in the left region with $\overset{\rightharpoonup}{m}_L=(m_{Lx},m_{Ly},m_{Lz})=m_L(\sin\theta\cos\beta,\sin\theta\sin\beta,\cos\theta)$, and fixed along the
$z$ axis perpendicular to the TI surface in the right region with $\overset{\rightharpoonup}{m}_R=(0,0,m_{Rz})$.
We can use a soft magnetic insulator for the left ferromagnet, which is controlled by a weak external magnetic field,
and a magnetic insulator with very strong easy-axis anisotropy for the right ferromagnet. The configuration between the left and right ferromagnets directly depends on the weak external magnetic field, where the
interlayer (RKKY) exchange coupling between left and right ferromagnets\cite{I.Garate2010} is ignored for the simplicity. In the middle segment,
there is no magnetization, but instead, a gate voltage $V_{0}$ is exerted.

Solving Eq.(\ref{Hamiltonian}), we obtain the wave function in the left region as following:
\begin{eqnarray}
\psi _{L}(x\leq 0)=A\left(
\begin{array}{c}
\frac{\upsilon _{F}\hbar k_{x}+m_{Lx}-i(\upsilon _{F}\hbar k_{y}+m_{Ly})}{%
E-m_{Lz}} \\
1%
\end{array}
\right) e^{ik_{x}x}  \notag \\
+B\left(
\begin{array}{c}
\frac{-(\upsilon _{F}\hbar k_{x}+m_{Lx})-i(\upsilon _{F}\hbar k_{y}+m_{Ly})}{%
E-m_{Lz}} \\
1%
\end{array}
\right) e^{-i(k_{x}+\frac{2m_{Lx}}{\upsilon _{F}\hbar})x}
\end{eqnarray}
where the Fermi energy lies in the upper bands of Dirac cone, and $E>0$. We also define
$\phi$ as the incident angle, then $k_{x}=(\sqrt{E^{2}-m_{Lz}^{2}}\cos\phi-m_{Lx})/\upsilon _{F}\hbar $, $k_{y}=(\sqrt{
E^{2}-m_{Lz}^{2}}\sin\phi-m_{Ly})/\upsilon _{F}\hbar$. The wave function in normal region $
\psi_{C}$ depends on the gate voltage. If $V_0\neq E $,
\begin{eqnarray}
\psi _{C}(0 \leq x\leq L)&=&C\left(
\begin{array}{c}
\frac{\upsilon _{F}\hbar (k_{x}^{\prime }-i k_{y})}{E-V_{0}} \\
1%
\end{array}
\right) e^{ik_{x}^{\prime }x}  \notag \\
&+&D\left(
\begin{array}{c}
\frac{-\upsilon _{F}\hbar (k_{x}^{\prime }+i k_{y})}{E-V_{0}} \\
1%
\end{array}
\right) e^{-ik_{x}^{\prime }x}
\end{eqnarray}
where $k_{x}^{\prime }=\pm\sqrt{((E-V_{0})/\upsilon _{F}\hbar)^{2}-k_{y}^{2}}$ with the $\pm$
corresponding to the upper bands and the lower bands of the Dirac cone respectively, and if $%
V_{0}=E$,\cite{M.I.Katsnelson2006} it becomes
\begin{eqnarray}
\psi _{C}(0 \leq x\leq L)&=&C\left(
\begin{array}{c}
0 \\
1%
\end{array}
\right) e^{-k_{y}x}  \notag \\
&+&D\left(
\begin{array}{c}
1 \\
0%
\end{array}
\right) e^{k_{y}x}
\end{eqnarray}
The wave function in the right region is:
\begin{eqnarray}
\psi _{R}(L\leq x)=F\left(
\begin{array}{c}
\frac{\upsilon _{F}\hbar (k_{x}^{\prime \prime }-i k_{y})}{E-m_{Rz}} \\
1%
\end{array}
\right) e^{ik_{x}^{\prime \prime }x}
\end{eqnarray}
with $k_{x}^{\prime \prime }=\sqrt{(E^{2}-m_{Rz}^{2})/(\upsilon _{F}\hbar)^{2}-k_{y}^{2}}$. There
exists a translation invariance along the y direction, so the momentum $k_{y}$ is
conserved in the three regions, and we omit the part $e^{ik_{y}y}$ in wave functions. These piecewise wave functions are connected
by the boundary conditions:
\begin{equation}
\psi _{L}(0)=\psi_{C}(0), \hspace{0.5cm} \psi_{C}(L)=\psi_{R}(L)
\end{equation}
which determine the coefficients A,B,C,D and F in the wave functions.

As a result, according to the Landauer-B\"{u}ttiker formula \cite{S.Datta1995}, it is
straightforward to obtain the ballistic conductance $G$ at zero temperature
\begin{eqnarray}
G =\frac{e^{2}w_{y}}{h\pi}\frac{E_{F}}{\upsilon _{F}\hbar}\frac{1}{2}\int_{-\frac{\pi }{2}}^{%
\frac{ \pi }{2}}d\phi \frac{F^{\ast }F}{A^{\ast }A}\frac{%
(E_{F}-m_{Lz})\upsilon _{F}\hbar k_{x}^{\prime \prime }}{(E_{F}-m_{Rz})E_{F}}
\end{eqnarray}
where $w_{y}$ is the width of interface along the y direction, which is much larger than $L$, and we take $E$ as $E_F$, because in our case the electron transport happens around the Fermi level.

\section{Numerical Results and Discussions}

We focus on the two cases about the electronic transport controlled by a gate
voltage through this 2D topological ferromagnet/normal/ferromagnet junction.
One is the conductance $G$ and the magnetoresistance when the
magnetizations in the left and right FM are collinear in the z-direction, and another is
the influence of the magnetization component along the x/y direction on the
conductance.

\subsection{The conductance and MR for collinear magnetization}

\begin{figure}[ht]
%\begin{center}
\includegraphics[width=0.51\textwidth]{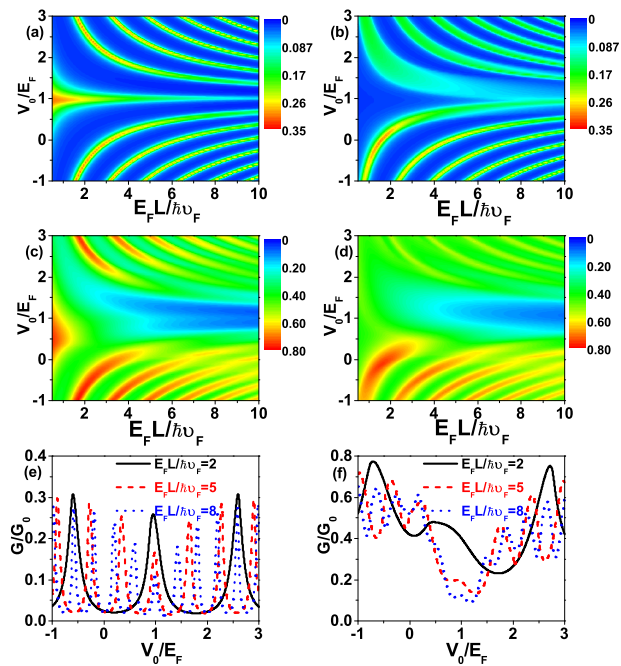}
\caption{(Color online) The normalized conductance $G/G_{0}$ as a function of
$E_{F}L/\upsilon _{F}\hbar$ and $V_{0}/E_{F}$, $m_{Lz}=m_{Rz}=0.95E_{F}$ in (a) and (b),
$m_{Lz}=m_{Rz}=0.6E_{F}$ in (c) and (d). (a) and (c) correspond to
the parallel configuration, (b) and (d) correspond to the antiparallel
configuration. (e) and (f) are the sections of (a) and (c) for three
$E_{F}L/\hbar\upsilon _{F}$'s  respectively.}
\label{fig2}
\end{figure}

We show the normalized conductance $G/G_{0}$ as a function of $k_{F}L$ and $%
V_{0}/E_{F}$ of parallel (Fig. a and c) and antiparallel (Fig.
b and d) configurations for two different magnetizations along the z-axis in Fig. \ref{fig2},
where $G_{0}=\frac{e^{2}w_{y}}{h\pi }\frac{E_{F}}{\upsilon _{F}\hbar}$. In Figs. \ref{fig2}(a) and \ref{fig2}(b) we choose $%
m_{Lz}=m_{Rz}=0.95E_{F}$, while in Figs. \ref{fig2}(c) and \ref{fig2}(d) $m_{Lz}=m_{Rz}=0.6E_{F}$.
In Fig. \ref{fig2}(a) the gap of surface state in the left and right ferromagnet regions
opened by the magnetization along the z-axis is $0.95E_{F}$. The conductance
oscillates with the gate voltage $V_{0}$ (parameters $E_{F}L/\hbar\upsilon _{F}$
and
$V_{0}/E_{F}$ in Fig. \ref{fig2} are dimensionless). The maximum of conductance
gradually decreases as the width increases. The minimum of conductance can
approach to zero. The change of conductance between maximum and minimum by gate voltage
is similar to the spin field-effect transistor, in which the conductance
modulation arises from the spin precession due to the spin-orbit coupling\cite{S.Datta1990}. The
gate voltage can be used to change the $k_{x}^{\prime}$ such that the
phase factor $k_{x}^{\prime}L$ of quantum interference in the normal
segment can be changed. The oscillation period of conductance with
respect to $V_{0}$ depends on the width $L$ and decreases with the increase
of width $L$. The conductance has a period $\pi$ with respect to
$z=V_{0}L$, when $V_{0}\rightarrow \infty, L\rightarrow 0$, in 2D
topological ferromagnet/ferromagnet junction\cite{T.Yokoyama2010,B.Soodchomshom2010}.

In Fig. \ref{fig2}(b), the conductance changes with the width $L$ and gate voltage $V_{0}$
in the same way as in Fig. \ref{fig2}(a). The difference is that the conductance is
maximum in Fig. \ref{fig2}(b) while it is minimum in Fig. \ref{fig2}(a) and vice versa. The
conductance in Fig. \ref{fig2}(c) and Fig. \ref{fig2}(d) show the same variation tendency with
the width $L$ and gate voltage $V_{0}$ as Fig. \ref{fig2}(a) and Fig. \ref{fig2}(b), respectively.
However both the maximum and minimum of conductance in Fig. 2(c) and Fig. 2(d)
are larger than those in Fig. \ref{fig2}(a) and Fig. \ref{fig2}(b),
since the gap of surface states in left and right
ferromagnet regions is $0.6E_{F}$ in Fig. \ref{fig2}(c) and Fig. \ref{fig2}(d). The conductance changes more
obviously with the gate voltage at the side of $V_{0}/E_{F}<1$ than at the side of $V_{0}/E_{F}>1$.
In Fig. \ref{fig2}, both the maximum and minimum of the conductance
become smaller when the gate voltage is closer to the Fermi energy,
because the number of the incident wave functions transported through the normal segment by the evanescent waves (imaginary $k^\prime_x$) becomes bigger.
Fig. \ref{fig2} shows that the conductance of this 2D topological
ferromagnet/normal/ferromagnet junction could be changed by the same way as
that in the spin field-effect transistor. While for the reason of the angular
spectrum of electrons in the surface plane and the linear dispersion
relation, how to get a large maximum/minimum ratio of the conductance is
important for a transistor.

\begin{figure}[tbph]
%\centering
\includegraphics[width=0.5\textwidth]{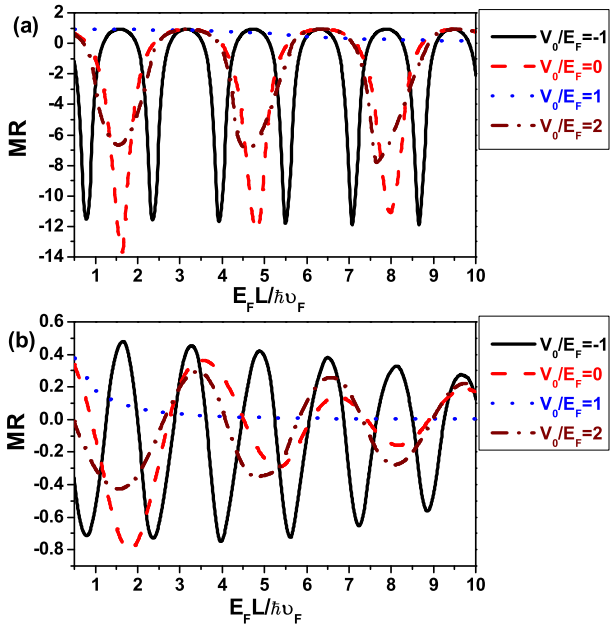}
\caption{(Color online) The MR as a function of the width
$E_{F}L/\hbar\upsilon _{F}$
with different gate voltage $V_{0}$. (a) $m_{Lz}=m_{Rz}=0.95E_{F}$;
and (b) $ m_{Lz}=m_{Rz}=0.6E_{F}$.}
\label{fig3}
\end{figure}

After obtaining the conductance $G_{P}$ of parallel configuration and $G_{AP}$ of antiparallel
configuration, we can get the MR directly, which is defined as $
MR=(G_{P}-G_{AP})/G_{P}$. Compared with the conductance in Fig. \ref{fig2}(a) and Fig. \ref{fig2}(c),
the conductance in Fig. \ref{fig2}(b) and Fig. \ref{fig2}(d) shows a property indicated below. On the one hand , the conductance in
the antiparallel configuration can be less than that in the parallel configuration
as in the conventional spin valve\cite{I.Zutic2004,S.A.Wolf2001,A.Fert2008}
and its counterpart in graphene\cite{C.Bai2008}. On the other hand,
the conductance in the antiparallel configuration
can also be larger than that in the parallel configuration, which is an anomalous
electronic transport property of topological spin-valve junction. Fig. \ref{fig3}
shows the MR as a function of the width $L$. When $V_{0}/E_{F}\neq 1$, the MR oscillates with
the width $L$. The amplitude and period of oscillation of MR depend on the
gate voltage $V_{0}$. When $V_{0}/E_{F}=1$, the MR does not oscillate and
decreases monotonically with the increase of $L$, because the Fermi surface
of normal segment is at the Dirac point in this case and the corresponding
density of states is zero while the conductance is not zero, which is a
typical property of Dirac fermion system\cite{M.I.Katsnelson2006}. The MR could be negative
for the anomalous electronic transport\cite{T.Yokoyama2010,T.Yokoyama2011}.
The maximum of MR in Fig. \ref{fig3}(a) is
larger than that in Fig. \ref{fig3}(b), and it can approach $100\%$. The big negative
MR (more than -10) in Fig. \ref{fig3}(a) also means a big variation of conductance
between parallel and antiparallel configuration.

Next we will discuss the underlying physics quantitatively to
understand the above results clearly. Since the electrons from all incident
angles give contributions to the conductance which is proportional to the electron
transmission probability, the physical origin of conductance
oscillating with the width $L$ and gate voltage $V_{0}$ in Fig. \ref{fig2} is a direct
result of summation of electron transmission probability over all incident
angles.

\begin{figure}[th]
%\centering
\includegraphics[width=0.51\textwidth]{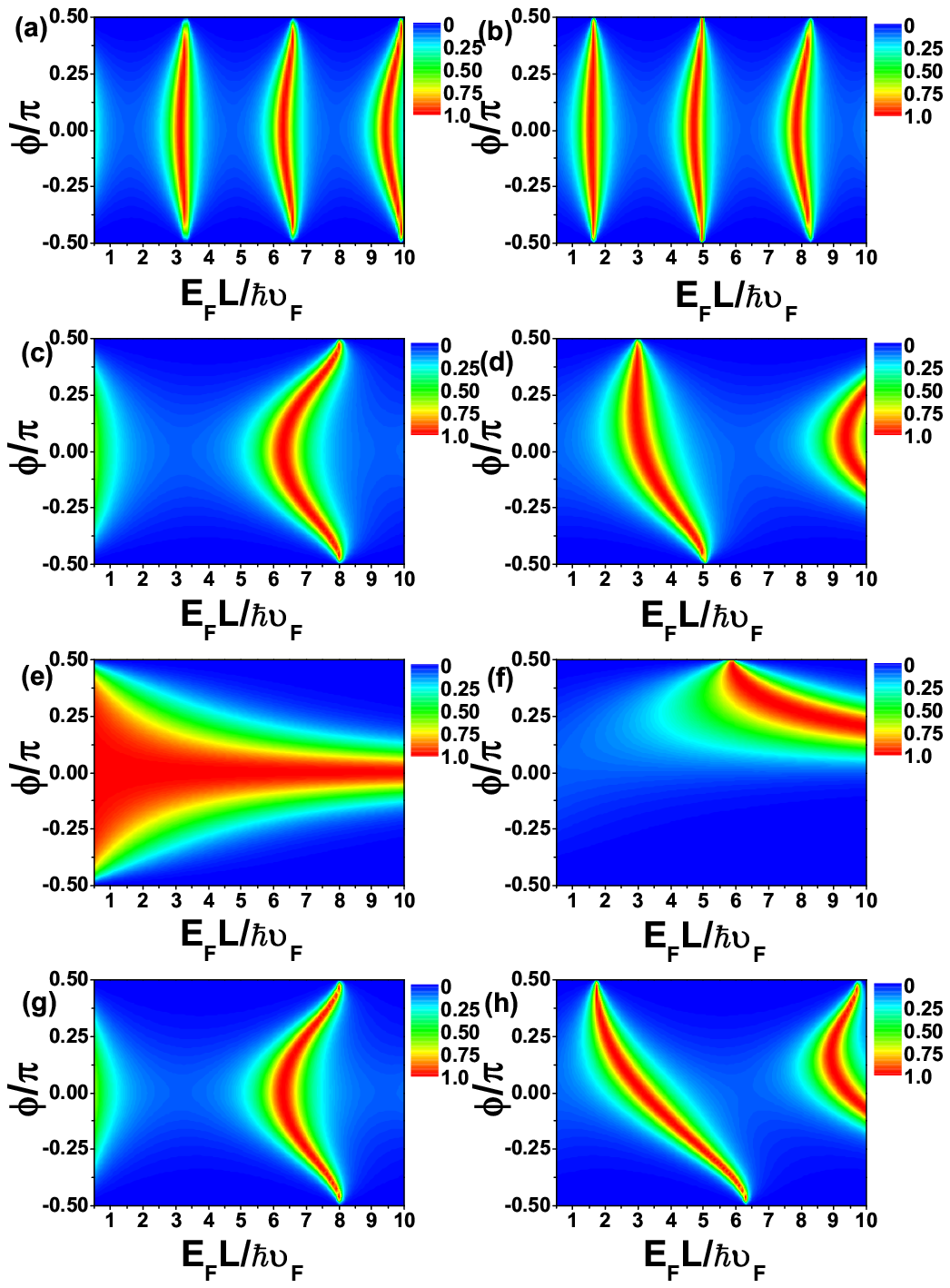}
\caption{(Color online) Transmission probability as a function of
incident angle $\phi$ and width $E_{F}L/\hbar\upsilon _{F}$
where $m_{Lz}=m_{Rz}=0.95E_{F}$, we choose the parallel configuration at
the left-hand side and the antiparallel configuration at the
right-hand side, and the gate voltage $V_{0}/E_{F}$ in (a) and (b),
(c) and (d), (e) and (f), (g) and (h) are 0, 0.5, 1 and 1.5,
respectively.}
\label{fig4}
\end{figure}
\begin{figure}[th]
\centering
\includegraphics[width=0.4\textwidth]{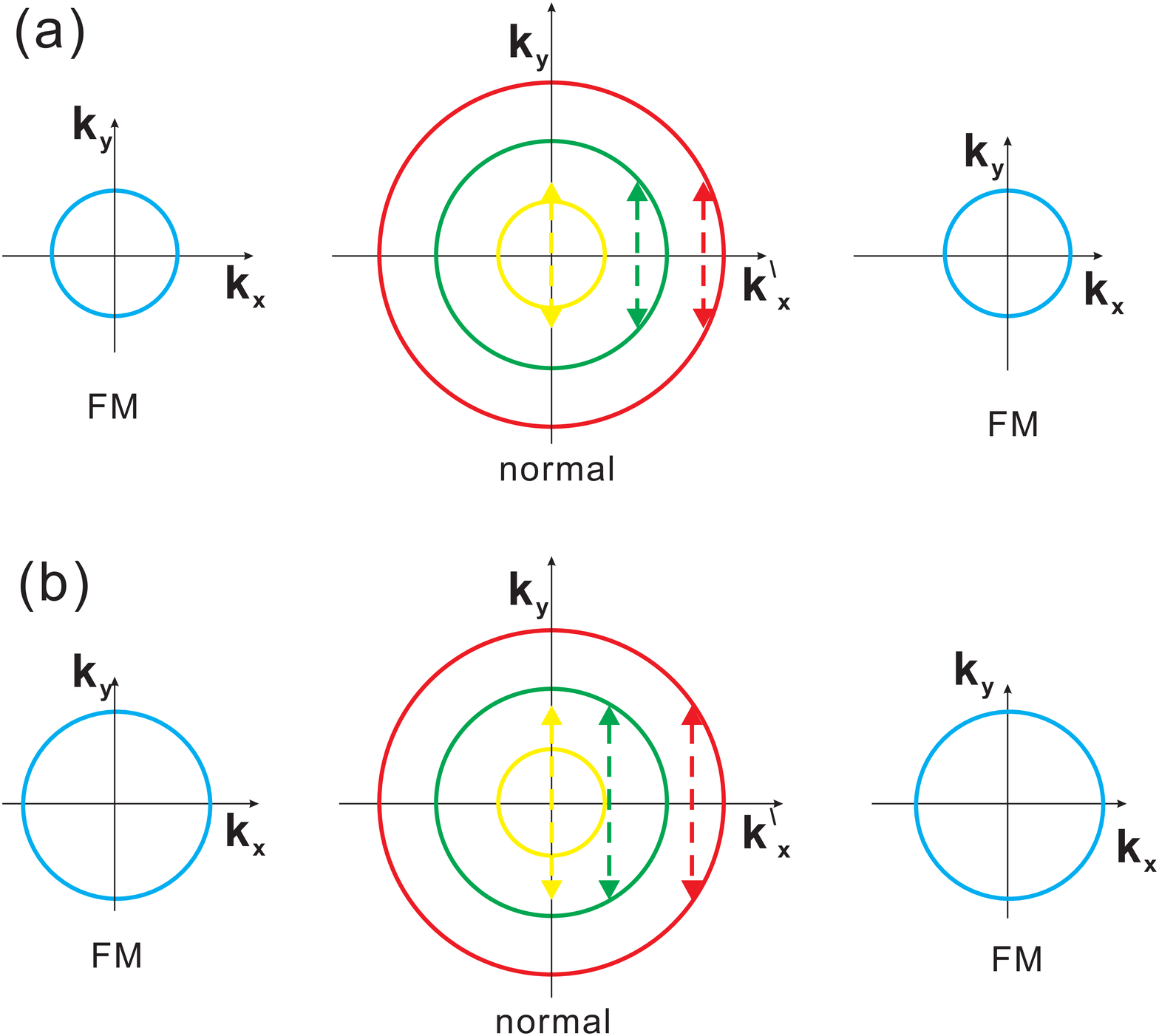}
\caption{(Color online) Fermi surfaces of the
ferromagnet/normal/ferromagnet junction in momentum space, where the different colored Fermi surfaces
in the normal segment stand for the cases with different gate voltages and the dashed
lines have the same length which equals to the range of momentum $k_y$ of incident wave function in (a) and (b), respectively.
(a) $m_{Lz}=m_{Rz}=0.95E_{F}$, and (b) $m_{Lz}=m_{Rz}=0.6E_{F}$.}
\label{fig5}
\end{figure}

Fig. \ref{fig4} plots the transmission probability as a function of incident angle
$\phi$ and width $L$ for different gate voltage $V_{0}$. We find that the
transmission probability mainly oscillates with the width $L$. Its period of
oscillation becomes large as the gate voltage increases from $V_{0}/E_{F}=0$ to
$V_{0}/E_{F}=1$. The reason for such a change can be illustrated in
Fig. \ref{fig5}. Because the wave functions in the left and right FMs are connected
through the wave function in normal segment, the transmission probability depends
on the phase factor $k_{x}^{\prime}L$. Due to the conservation of momentum
$k_{y}$, $k_{x}^{\prime}$ depends on the gate voltage. When the gate voltage
from $V_{0}/E_{F}=0$ to $1$, the Fermi surface for the normal region reduces as in Fig. \ref{fig5},
and $k_{x}^{\prime}$ reduces too, such that the transmission probability has a
longer periodicity with the width $L$ and changes considerably  with incident angles
as shown in Fig. \ref{fig4}(a) or \ref{fig4}(b) and \ref{fig4}(c) or \ref{fig4}(d). In these cases, the electronic transport
through the normal segment occurs in the upper bands of Dirac cone.
Although the Fermi surface for the normal segment in
Fig. \ref{fig4}(g) or \ref{fig4}(h) is equal to that in Fig. \ref{fig4}(c) or \ref{fig4}(d), their transmission probability
is different, because in Fig. \ref{fig4}(g) or \ref{fig4}(h) the electronic transport through the normal
segment occurs in lower bands of Dirac cone.
When the gate voltage $V_{0}/E_{F}=1$, the electronic
transport through the normal segment is totally due to the evanescent waves, the
transmission probability is not a periodic function of width $L$ as in Fig.
\ref{fig4}(e) or \ref{fig4}(f).

Now we consider the influence of magnetization configuration on the
transmission probability. It is clearly that the transmission probability is
an even function of the incident angle $\phi$ in the parallel configuration at
the left-hand side of Fig. \ref{fig4}, while it is not an even function of the
incident angle $\phi$ in the antiparallel configuration at the right-hand side.
This is unusual, because the transmission probability is an even function of
the incident angle $\phi$ on the antiparallel configuration in its counterpart
in graphene\cite{C.Bai2008}.
This unusual property arises from the unequal spinor parts of the incident and transmission wave functions. At the normal incidence ($\phi=0$), the period of the transmission
probability with the width L in the parallel configuration is the same as that in
the antiparallel configuration and the position of maximum of the transmission
probability has a shift of the half-period between two configurations. Now with
the help of Figs. \ref{fig4} and \ref{fig5}, the properties of conductance in Fig. \ref{fig2}(a) and \ref{fig2}(b)
and MR in Fig. \ref{fig3}(a) could be understood explicitly.

\begin{figure}[th]
\centering
\includegraphics[width=0.5\textwidth]{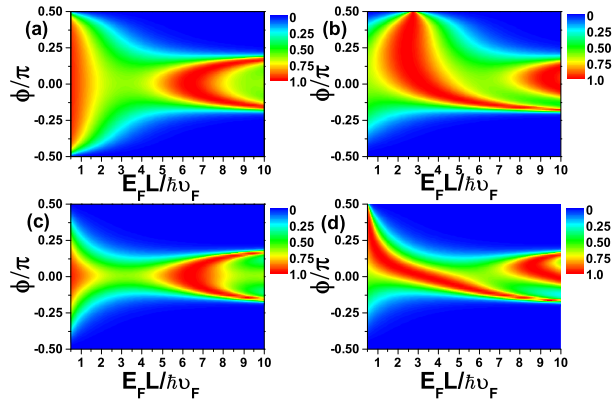}
\caption{(Color online) Transmission probability as a function of incident angle
$\phi$ and the width $E_{F}L/\hbar\upsilon _{F}$, where $m_{Lz}=m_{Rz}=0.6E_{F}$,
and we choose the parallel configuration at the left-hand side and the antiparallel
configuration at the right-hand side, and the gate voltage $V_{0}/E_{F}=$ 0.5
and 1.5 in (a) and (b), (c) and (d), respectively.}
\label{fig6}
\end{figure}

When the magnetizations in the left and right FMs are taken as $0.6E_{F}$ in
Fig. \ref{fig5}(b), one may see that the gaps of the surface states in the left and right ferromagnet regions decrease,
and the Fermi surfaces in the left and right FMs become large.
So, the range of $k_{y}$ expands, and those of
$k_{x}^{\prime}$ and the phase factor $k_{x}^{\prime}L$ expand too. The
transmission probability in Fig. \ref{fig6} changes more dramatically than in Figs. \ref{fig4}(c) and \ref{fig4}(d),
\ref{fig4}(g) and \ref{fig4}(h). Therefore, as the gap of surface states in left and right ferromagnet
regions decreases, more incident electronic states
will contribute to the conductance, such that the conductance becomes larger
on the whole, and more unsymmetrical about the gate voltage $V_{0}/E_{F}=1.0$
in Fig. \ref{fig2}(c) and Fig. \ref{fig2}(d). The MR in Fig. \ref{fig3}(b) could be understood, similarly.

\subsection{The influence of x/y component of magnetization on the
conductance}

Now we consider the influence of x/y component of magnetization on the
conductance. First, we choose the z component of magnetization in the left and
right FM to be equal as that in subsection A. We find that the influences of
x/y component of magnetization on the conductance are quite different. The x
component of magnetization has no influence on the conductance, while the y
component of magnetization has a great influence on the conductance. Because the x
component of magnetization just moves the Fermi surface along the x axis, the
states contributing to the conductance do not change, while the y component of
magnetization shifts the Fermi surface in the left FM along the y direction and
decreases the number of incident electron states that contribute to the
conductance.
\begin{figure}[th]
%\centering
\includegraphics[width=0.45\textwidth]{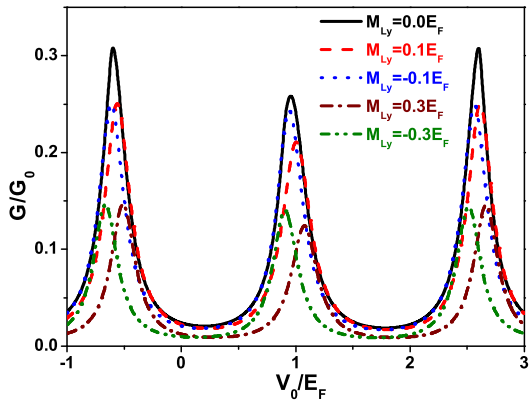}
\caption{(Color online) The conductance as a function of the gate voltage $V_{0}$ for different $m_{Ly}$, where
$E_{F}L/\hbar\upsilon _{F}=2$ and $m_{Lz}=m_{Rz}=0.95E_{F}$.}
\label{fig7}
\end{figure}
The influence of $m_{Ly}$ on the conductance is shown in Fig. \ref{fig7}. It is seen that
the conductance decreases with increasing $|m_{Ly}|$, so a large $|m_{Ly}|$ can
lead the conductance to be zero. We also discover that the influence of
magnetization $m_{Ly}$ on the conductance is different from that of $-m_{Ly}$.

Second, by keeping the magnetizations in the left and right
FMs the same value, the direction of magnetization in the left FM is changed in the x-z plane ($\beta=0$)
or in the y-z plane ($\beta=\pi/2$), where $\theta$ and $\beta$ are indicated as shown in Fig. 1.
\begin{figure}[th]
%\centering
\includegraphics[width=0.51\textwidth]{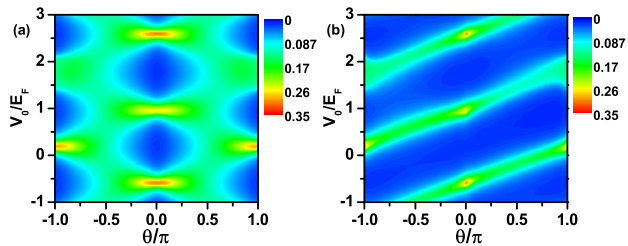}
\caption{(Color online) The conductance as a function of $\theta$
and gate voltage $V_{0}/E_{F}$ for $E_{F}L/\hbar\upsilon _{F}=2$, $
m=|(m_{Lx},m_{Ly},m_{Lz})|=|(0,0,m_{Rz})|=0.95E_{F}$, the angle $
\theta $ is (a) in the x-z plane ($\beta=0$)
and (b) in the y-z plane ($\beta=\pi/2$).}
\label{fig8}
\end{figure}
The conductance as a function of $\theta$ and the gate voltage $V_{0}$ is plotted
in Fig. \ref{fig8}, which is different from that in
ferromagnetic/normal/ferromagnetic graphene junction\cite{T.Yokoyama2011}.
The distinction between Figs. \ref{fig8}(a) and \ref{fig8}(b) is more obvious at $\theta =\pm 0.5\pi $,
where the conductance changes slightly with the gate voltage in Fig. \ref{fig8}(a)
while the conductance changes remarkably in Fig. \ref{fig8}(b). These
results are from different connections of wave functions between left and
right FMs. Since when $\theta =\pm 0.5\pi$, the spin in the right FM is parallel to
$(\upsilon _{F}\hbar k_{x}^{R},\upsilon _{F}\hbar k_{y},m)^{t}$,
\cite{T.Yokoyama2010}
and the spin in the left FM is parallel to
$(\upsilon _{F}\hbar k_{x1}\pm m,\upsilon _{F}\hbar k_{y1},0)^{t}$ in Fig. 8(a)
which satisfies the relation $E=\sqrt{%
(\upsilon _{F}\hbar k_{x1}\pm m)^{2}+(\upsilon _{F}\hbar k_{y1})^{2}}$,
while the spin in the left FM is parallel to
$(\upsilon _{F}\hbar k_{x2},\upsilon _{F}\hbar k_{y2}\pm m,0)^{t}$ in Fig. 8(b)
which satisfies the relation
$E=\sqrt{(\upsilon _{F}\hbar k_{x2})^{2}+(\upsilon _{F}\hbar k_{y2}\pm m)^{2}}$.
In this case, the z component of spin in the left FM is
$0$ in Figs. \ref{fig8}(a) and \ref{fig8}(b). Because in Fig. \ref{fig8}(b) the Fermi surface
of left FM shifts along the y direction about $\pm m$, the
difference of x component of spin between the left FM and right FM
in Fig. \ref{fig8}(a) is larger than that in Fig. \ref{fig8}(b).

Finally, we discuss the realization of our model. The bulk band gap of
topological insulator is small and depends on the materials, which is for
example, about $300$ meV in $Bi_{2}Se_{3}$, $100$ meV in $Bi_{2}Te_{3}$
\cite{H.Zhang2009,Y.Xia2009,Y.L.Chen2010},
and $22$ meV in HgTe\cite{C.Brune2011}. Far away from the Dirac point, the
surface electronic states exhibit large deviations from the simple Dirac cone in
$Bi_{2}Te_{3}$\cite{Y.L.Chen2009}.
The gap of surface states could be induced by putting the
magnetic insulator on the surface of a topological insulator (such as EuO, EuS and MnSe).
Depending on the interface match of the topological insulator and ferromagnetic
insulator, the gap is several to dozens of meV \cite{T.Yokoyama2010,H.Haugen2008,I.Vobornik2011,
Weidong Luo2012}.
The gate electrode could be attached to the topological insulator to
control the surface potential\cite{H.Steinberg2011,Y.Wang2012,J.R.Williams2007}.
The predicted properties of our model may be observed
when the Fermi energy of surface states is about $10-100$ meV, and
the junction width is about $10-100$ nm. The calculated results in
this paper are based on the ballistic transport. In order to observe
experimentally our predicted properties, a clean 2D topological
surface states with enough long mean free path is needed. It is
interesting to note that the surface of topological insulator with
such a long mean free path can be realized in experiments\cite{Y.Wang2012}.

\section{Summary}

In summary, we have studied the electronic transport properties of
the ferromagnet/normal/ferromagnet junction on the surface of a
strong topological insulator, where a gate voltage is exerted on the
normal segment with a finite width. It is found that the conductance
oscillates with the width of normal segment and the gate voltage.
The maximum of conductance gradually decreases as the width
increases and the minimum of conductance approaches zero. This
gate-controlled conductance behaves
in the same way as the spin
field-effect transistor does, but a further study is needed to
increase the maximum/minimum ratio of the conductance. The
magnetoresistance can be very large and could also be negative owing
to the anomalous transport. In addition, when there exists a
magnetization component in the 2D plane, it is shown that only the
magnetization component parallel to the junction interface has an
influence on the conductance.

\acknowledgments
One of authors (KHZ) acknowledges discussions with
Fei Ye and Zhe Zhang. This work is supported in part by the NSFC
(Grant Nos. 90922033, 10934008, and 10974253), the MOST of China
(Grant No. 2012CB932900 and 2013CB933401) and the CAS.

\end{document}